\documentclass{stsci_report}

\usepackage{multirow}
\usepackage{array}
\usepackage{graphicx}
\usepackage{siunitx}
\usepackage{pdfpages}
\usepackage{longtable}
\usepackage{booktabs}
\usepackage{float}
\usepackage{indentfirst}
\usepackage{caption}
\usepackage{hyperref}
\usepackage{multirow}
\usepackage{booktabs} 
\usepackage{xurl}
\bibliography{bibliography.bib}

\copyrighttext{Copyright\copyright\ \the\year\ The Association of Universities for Research in Astronomy, Inc. All Rights Reserved.}

\presubtitle{Instrument Science Report ACS 2025-01}
\title{The Effect of ``Roll-Drift'' in ACS/WFC Images}
\author{Yotam Cohen}
\date{April 25, 2025}

\begin{document}

\maketitle

\abstract{In 2024, due to some operational changes, the Hubble Space Telescope began exhibiting undeclared loss of lock events. This loss of lock can result in the smearing out of light from the target field during an exposure, which leads to data degradation, which in turn may require data to be retaken. In this work, we investigate this `roll-drift' effect in ACS/WFC images. We quantify the impact of roll-drift on measurable parameters in data by using simulations and existing data reduction techniques. We identify a threshold of one such measurable parameter beyond which data may likely be affected by roll-drift, so that users can quickly and easily assess whether their data needs further attention.}

\section{Introduction}

In June 2024, the Hubble Space Telescope (HST) transitioned to Reduced Gyro Mode (RGM), which has caused a number of changes in how the telescope operates.
Interested readers can find more information about RGM at the following page and the links therein: \url{https://hst-docs.stsci.edu/hsp/hst-general-science-policies/reduced-gyro-mode-tips-and-resources}.

Occasionally, during an exposure, HST experiences what is called a Loss of Lock (LoL) event, which causes the precise guiding to be lost, resulting in the pointing drifting slightly during the integration.
This drifting can cause the light from the target field to be smeared out in the resulting image, reducing the quality and usability of the data.
One of the unforeseen consequences of the RGM is an increased likelihood for these LoL events to occur.
We will henceforth refer to the collective phenomenon of HST drifting during an exposure resulting in smeared images as ``roll-drift."

Normally, these LoL events are detected by the observatory systems and are flagged as such in the exposure data.
However, sometimes, these events are not automatically detected, resulting in what is called an ``undeclared loss of lock event''.
This is very important because it creates an immediate need for new data to be monitored as it is made available, to check for possible signs of a LoL affecting the data quality.
Ever since transitioning to RGM, it appears that these undeclared LoL events have become more frequent.
Shortly after this started occurring, the ACS team began investigating the impact of this roll-drift effect on ACS data, as well as trying to automatically flag it when it does occur.
This report documents our work on those efforts.

\section{Analysis and Results}

\subsection{Roll-Drift Simulations}

\begin{figure}[h]
\centering
\includegraphics[width=\columnwidth]{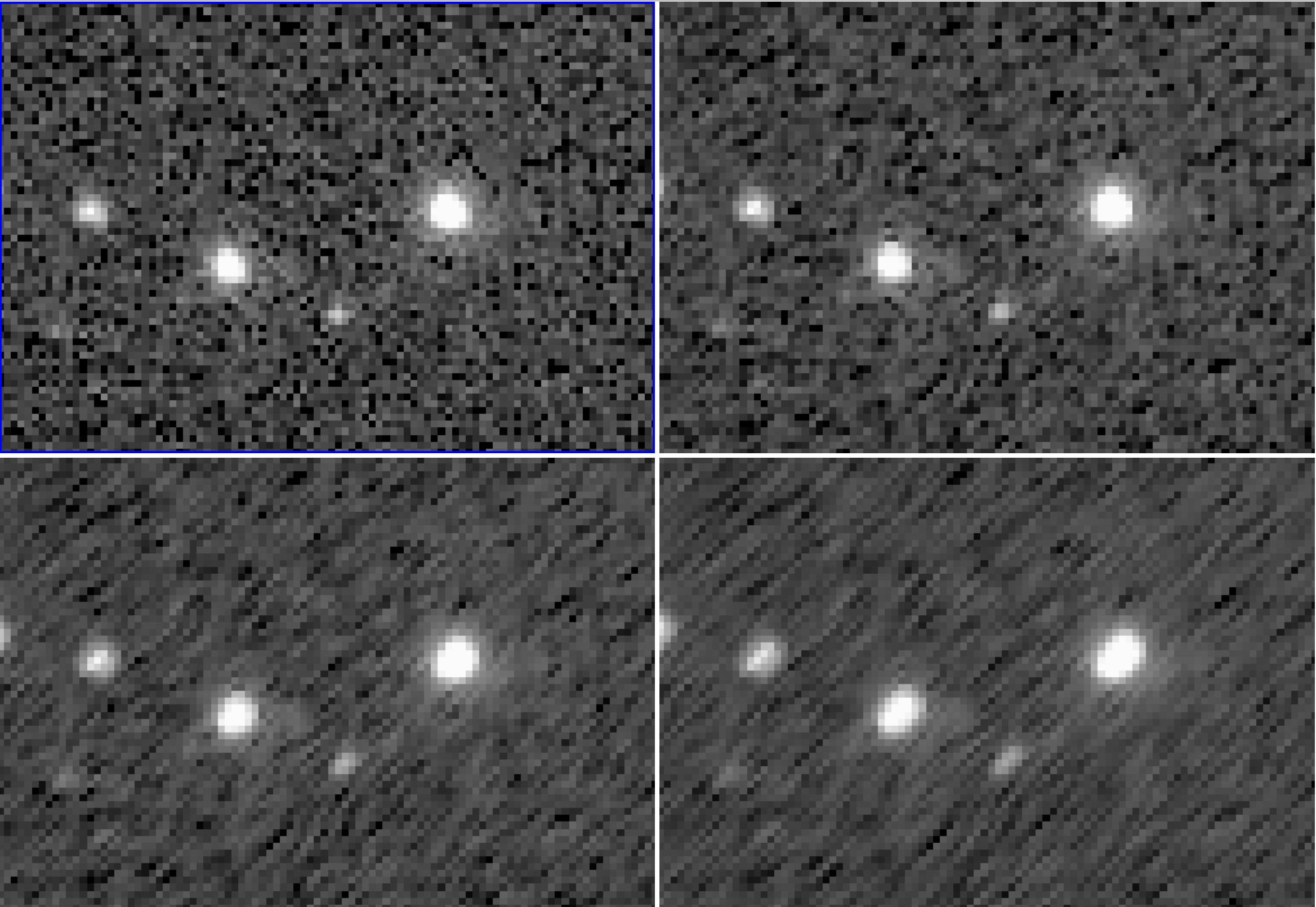}
\caption{Simulated roll-drift on a real ACS/WFC image, exposure ID \texttt{jf7921ulq}. Each panel shows the same cutout section of the image, centered on a few stars, in order to highlight the effect of the roll-drift. Top left: original image; no simulated roll-drift applied. Top right: 1 pixel of roll-drift. Bottom left: 2 pixels of roll-drift. Bottom right: 3 pixels of roll-drift. The color scale in each panel is identical.}
\label{fig:rolldrift}
\end{figure}

Our goal is to understand how roll-drift affects measurable parameters of ACS/WFC data, so that we can attempt to automatically flag exposures potentially affected by it.
In order to accomplish this, we have opted for an approach in which we simulate the effect of roll-drift on exposures, and then perform the data reduction steps that we would on real data.
There are several conceivable ways to simulate the effect of roll-drift on images.
For the purposes of this work, we use the following, relatively simple method:
We start with a real archival ACS/WFC \texttt{flc} image that is unaffected by roll-drift.
We then specify a reasonably small ``delta" displacement (e.g. 0.1 pixels) by which to increment the roll-drift, as well as a ``final" displacement (e.g. 5 pixels), and then we use the \texttt{scipy.ndimage.shift} module to shift the original image by every integer multiple of the delta displacement until the final displacement is reached.
Every intermediate shifted image is saved.
The direction of the displacement does not matter for the results of our study, but we choose the vector to have equal components in the x and y directions for this work.

Next, in order to produce the roll-drifted image for each amount of total displacement, we take all the shifted images with displacements up to and including that total amount, and average them together.
In order words, the final roll-drifted image for a given displacement is just the average of all the images in the stack before and including it.
This way of simulating the roll-drift and the resulting ``smearing" of the light in the images is physically reasonable, while also being quite simple and computationally-cheap to perform.
An example of the resulting simulated roll-drifted images is shown in Fig.~\ref{fig:rolldrift}.
As seen in this figure, small roll-drifts even up to about 1 pixel are visually difficult to notice, at least from just looking at a static image (compared to e.g. blinking multiple images).

\subsection{PSF-fitting Photometry with hst1pass}

With the set of simulated roll-drifted images in hand, we then move on to the next step of the data reduction/analysis, which is PSF-fitting photometry.
For this step, we run \texttt{hst1pass} \parencite{2022acs..rept....2A} on all the simulated roll-drifted images as well as the original (zero roll-drift) image.
Importantly, we configure \texttt{hst1pass} to run with the focus level set as a free parameter for the code to calculate, and we provide the library PSF for the particular filter from the existing library.
Allowing the focus level to vary is important because it allows the code to try to fit a reasonable PSF model to stars potentially smeared by the roll-drift, allowing for a sort of ``best case scenario.''
Also, this is typically how \texttt{hst1pass} is often configured for PSF-fitting photometry, so we follow that convention to make our results more meaningful to compare with typical use cases.

\subsection{Resulting Diagnostics}

\begin{figure}[ht!]
\centering
\includegraphics[width=\columnwidth]{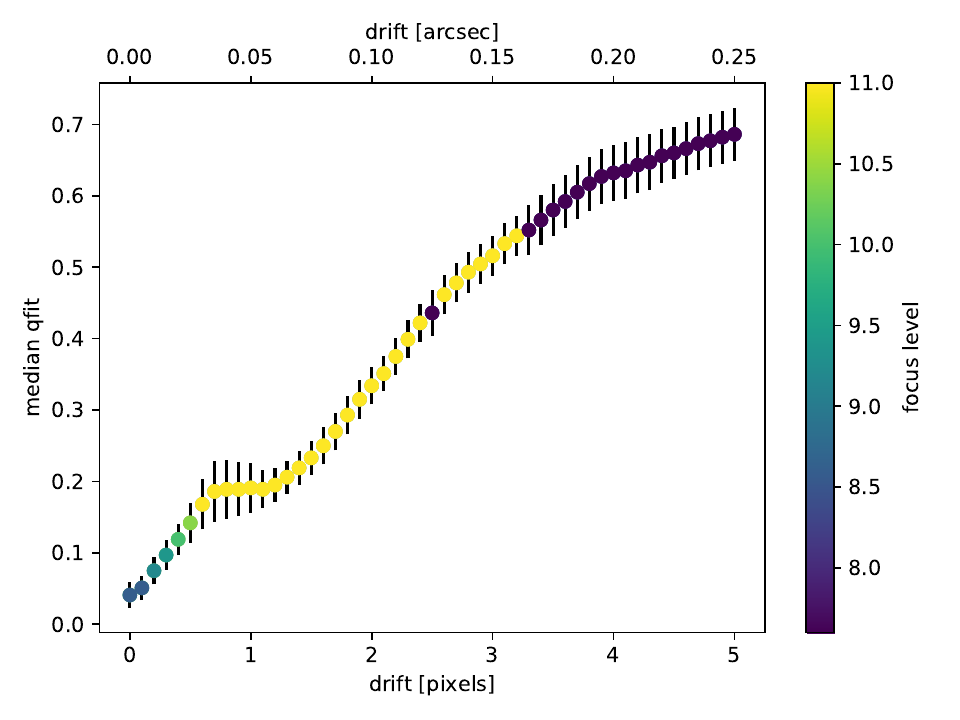}
\caption{Measured `qfit' value as a function of the amount of simulated roll-drift for ACS/WFC image with exposure ID \texttt{jf7921ulq}.
The `qfit' value is a measurement of how well the PSF model was fit to a given star, as explained in the \texttt{hst1pass} manual.
The values plotted on the y-axis are the sigma-clipped median `qfit' for all the stars in the image, with errorbars corresponding to the sigma-clipped standard deviation of the `qfit', and the color of the scatter points corresponding to the measured focus level of the image, as shown in the colorbar.
The x-axis shows the simulated roll drift in units of ACS/WFC pixels (bottom) and arcseconds (top).
The most important features of this plot are that the fitted focus level increases with increasing roll-drift, for small roll-drifts, and that beyond a roll-drift of about 1 pixel, the `qfit' values exceed a value of $\sim 0.2$, which is threshold for `good' stars.}
\label{fig:qfit_vs_drift}
\end{figure}

\begin{figure}[ht!]
\centering
\includegraphics[width=\columnwidth]{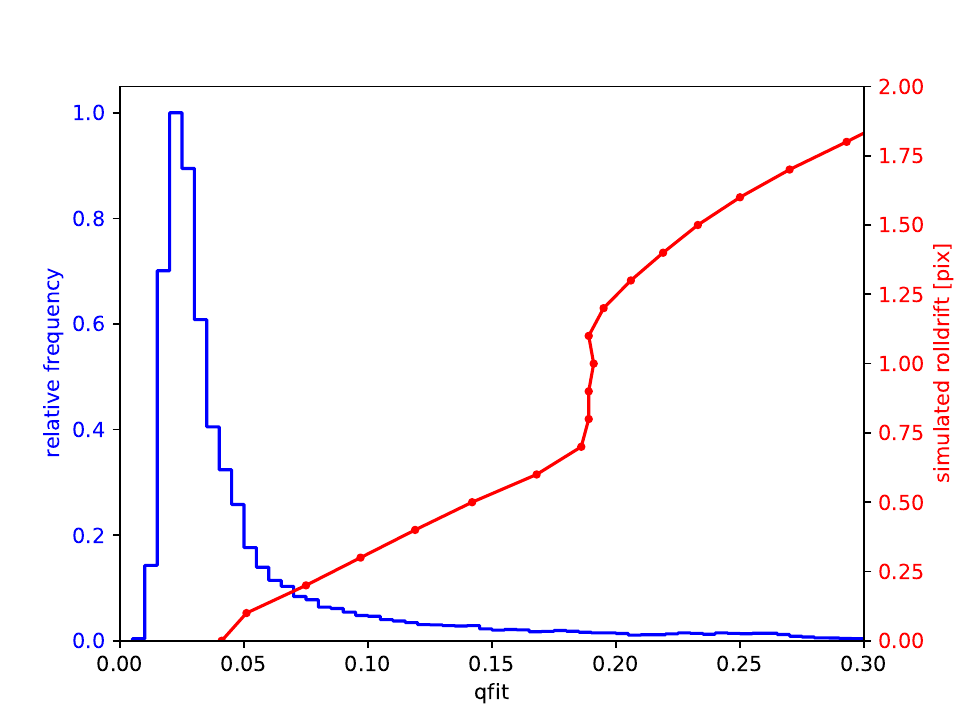}
\caption{The distribution of median `qfit' values for a large number of ACS/WFC images is plotted in blue. The curve from Fig.~\ref{fig:qfit_vs_drift} is plotted in red, with the axes flipped from that figure so as to be able to more easily compare with the `qfit' distribution.}
\label{fig:qfit_hist_with_drift}
\end{figure}

Once we have run \texttt{hst1pass} on all the images, we then move on to the analysis using the resulting catalogs.
The \texttt{hst1pass} code can calculate and output numerous parameters for each fitted source in a given image.
In this case, the parameter we are most interested in is called the `qfit` parameter (or just `q` for short), which is measurement of how well the focus-diverse PSF model fits to a given source \parencite{2022acs..rept....2A}.
For well-measured stars in exposures unaffected by roll-drift, their qfit value is expected to be very close to 0, typically around $\sim 0.02 - 0.04$.
According to the \texttt{hst1pass} manual \parencite{2022acs..rept....2A}, qfit of about 0.2 or more corresponds to poorly fit stars, and qfit values closer to 1.0 often correspond to resolved sources or cosmic rays.
For stars affected by roll-drift, both real and simulated, the smearing-out of the light from the star in the image will cause the PSF model to be more poorly fit, resulting in a higher value of qfit.
Our goals are then 1) to quantify the relationship between qfit and amount of roll-drift, and 2) to identify a threshold qfit beyond which we might like to automatically flag a given image as potentially affected by roll-drift.

In Fig.~\ref{fig:qfit_vs_drift}, we plot the median qfit for all the stars in the image, as a function of simulated roll-drift, for the example exposure we have been following in this report thus far.
The errorbars shown are the sigma-clipped standard deviation of the qfit values for each catalog, and the color of the scatter points corresponds to the calculated focus level, as shown in the colorbar.
We note a few important features of this plot.
First, we see that there is an overall increasing trend of qfit with increasing roll-drift, as expected -- the more smeared out the stars are, the more poorly the PSF model fit is.
We also notice that the calculated focus level increases steadily between 0 to about 0.6 pixels of roll-drift.
This also makes sense, as the code has to use more elongated PSF models \parencite{2018acs..rept....8B} to fit the more smeared out stars, which manifests as a higher focus level.
We then notice that the curve flattens out between about 0.7 to 1.1 pixels of roll-drift, and that the measured focus level here is the same.
This is where the PSF fitting is starting to break down, but that, interestingly, the elongated PSF models can fit stars equally well in that range of roll-drift.
Beyond about 1.1 pixels of roll-drift, the qfit continues to increase beyond the "good" threshold of $\sim 0.2$, and the code is unable to fit a PSF model well, meaning that PSF fitting photometry beyond this amount of roll-drift is fraught.
This plot is one of the major pieces of understanding how roll-drift affects the quality of image data and measurable parameters of the image.

The other piece we need for this understanding is the plot shown in Fig.~\ref{fig:qfit_hist_with_drift}.
There are two separate things plotted here.
In blue (and on the left y-axis) is a histogram of median qfit values for over 70,000 normal (i.e. not suffering from roll-drift) ACS/WFC archival images.
In red (and on the right y-axis) we plot the same curve shown in Fig.~\ref{fig:qfit_vs_drift}, but with the axes inverted so that we can make a meaningful comparison with the histogram plotted.
This plot has several important features.
First, we note that the distribution of qfit values of well-measured stars peaks at around 0.02, and that the distribution has a long, but suppressed, tail out to larger values of qfit.
The vast majority of the distribution lies between qfit values between about 0.0 and 0.075, as expected.
Next we look at the red curve and observe the following.
Zero roll-drift roughly coincides with a qfit close to the peak of the distribution, as expected.
Beyond that, however, up to a roll-drift amount of anywhere between about 0.1 to 0.4 pixels, the qfit of the drifted image is consistent with a range of qfit values that are within a reasonable range for unaffected images.
This is to say that, for images affected by small but non-zero roll-drift, the images may not exhibit any obvious visual indication of smearing/roll-drift, and the qfit value may appear as acceptable.
In other words, small amounts of roll-drift may be largely indistinguishable from zero roll-drift.
On one hand, this may be viewed positively, as images impacted by small amounts of roll-drift may still be entirely usable for some science cases.
On the other hand, this could also potentially adversely affect high precision scientific measurements with little or no indication as to the source of discrepancies.

\section{Recourse and Recommendation to Users}
In light of our findings in this work, the ACS team has set up an automated routine that runs \texttt{hst1pass} on all new ACS/WFC data as soon as it is taken, tabulates the qfit values, and sends members of the ACS team a daily email with that information.
The ACS team can then manually inspect any new exposures with a potentially suspect qfit value for additional signs of roll-drift or other data abnormalities.
Users of ACS/WFC data taken any time after HST entered RGM may also consider running \texttt{hst1pass} on their data, examining the qfit values, and visually inspecting the data.
The rate of occurrence of undeclared loss of lock and resulting roll-drift seems to be quite low, but if a user is unsure whether their data may be affected by roll-drift, they are encouraged to contact the ACS team via the STScI Help Desk (\url{https://stsci.service-now.com/hst}) for further assistance.

\section{Conclusion}
In this work, we performed an initial exploration of how loss of telescope guiding during an exposure can affect resulting ACS/WFC images.
We performed simulations of the smearing out of the resulting light during these events, and we analyzed measurable parameters of the resulting images in an effort to devise a way to automatically flag new data potentially impacted by this.
The ACS team then set up said flagging and is actively monitoring new data.

\section*{Acknowledgements}
The authors would like to acknowledge the entire ACS team for their support in this work.

\printbibliography

@MISC{2022acs..rept....2A,
       author = {{Anderson}, Jay},
        title = "{One-Pass HST Photometry with hst1pass}",
 howpublished = {Instrument Science Report ACS 2022-02},
         year = 2022,
        month = jul,
        pages = {2},
       adsurl = {https://ui.adsabs.harvard.edu/abs/2022acs..rept....2A}
}

@MISC{2018acs..rept....8B,
       author = {{Bellini}, Andrea and {Anderson}, Jay and {Grogin}, Norman A.},
        title = "{Focus-diverse, empirical PSF models for the ACS/WFC}",
 howpublished = {Instrument Science Report ACS 2018-8},
         year = 2018,
        month = nov,
        pages = {8-8},
       adsurl = {https://ui.adsabs.harvard.edu/abs/2018acs..rept....8B}
}

\end{document}